\documentclass[10pt,final,journal,letter,oneside,twocolumn]{IEEEtran}
\usepackage{cite}
\usepackage{ulem}
\usepackage{soul,color}
\usepackage{srcltx}
\usepackage{multirow}
\usepackage[tbtags,sumlimits,nointlimits,reqno]{amsmath}
\usepackage{graphicx,dblfloatfix}
\usepackage{float}
\usepackage{subfigure}
\usepackage{psfrag}
\usepackage{color}
\usepackage{amssymb}
\usepackage{makecell}
\usepackage{array}
\begin{document}

\title{Compact Reflection-Type Phaser Using Quarter-Wavelength Transmission Line Resonators}

\author{%
       Weiwei~Liao,~\IEEEmembership{Student~Member,~IEEE}
       Qingfeng~Zhang,~\IEEEmembership{Member,~IEEE}
       ,~Yifan~Chen,~\IEEEmembership{Senior~Member,~IEEE}
       and~C.~Caloz,~\IEEEmembership{Fellow,~IEEE}
\thanks{Manuscript received Nov. 2014.}
\thanks{Weiwei~Liao, Qingfeng~Zhang and Yifan~Chen are with the Department of Electrical and Electronic
Engineering, South University of Science and Technology of China, China, Email: zhang.qf@sustc.edu.cn.

C. Caloz is with the Department of Electrical Engineering, PolyGrames Research Center, \'{E}cole Polytechnique de Montr\'{e}al, Montr\'{e}al, QC, Canada H3T 1J4, and with King Abdulaziz Universitity, Saudi Arabia.}}
\markboth{IEEE Microwave and Wireless Component Letters}%
{Shell \MakeLowercase{\textit{et al.}}: Bare Demo of IEEEtran.cls for Journals}

\maketitle

\begin{abstract}
A compact reflection-type phaser composed of quarter-wavelength transmission line resonators interconnected by alternating K- and J-inverters is proposed. A design method is also presented. To validate this method, a $4^\text{th}$-order example is designed and fabricated. The proposed phaser is shown to exhibit the benefits of smaller size, easier fabrication and suppressed even-order harmonics compared with previously reported half-wavelength phasers.
\end{abstract}

\IEEEpeerreviewmaketitle
\section{Introduction}
\IEEEPARstart{R}{eal-time} analog signal processing (R-ASP), which consists in manipulating signals in their pristine analog and real-time form, exhibits several benefits such as high speed, low cost and low consumption, compared with digital signal processing (DSP) at microwave and millimeter-wave frequencies~\cite{2AnalogSignal}. Several R-ASP applications have been recently reported in~\cite{3CRLHDelayLinePulsePosition,4CompressiveReceiverUsingaCRLH-Based,5MicrowaveAnalogReal-TimeSpectrum,1NonuniformlyCoupledMicrostrip,7ChiplessRFIDSystemBasedonGroupDelay}.

The phaser, a component providing specifiable (non-flat) group delay versus frequency response, is the core element of an R-ASP system~\cite{2AnalogSignal}. It can be either of transmission type (2-port) or reflection type (1-port). Transmission-type phasers, implemented using allpass C-sections~\cite{6Group-DelayEngineeredNoncommensurate} or bandpass cross-coupled resonators~\cite{Zhang_TMTT_cross}, are typically complicated to synthesize. Reflection-type phasers feature simpler design, at the expense of requiring an external circulator or hybrid coupler to transform into a 2-port component. A complete synthesis technique for reflection-type phasers, based on half-wavelength transmission line resonators and K-inverters, was reported in~\cite{12SynthesisofNarrowbandReflection}.

This paper introduces a reflection-type phaser, based on \emph{quarter-wavelength} transmission line resonators and alternating K/J-inverters as quarter-wavelength filters~\cite{10DIRECT-COUPLED}. Compared with that in~\cite{12SynthesisofNarrowbandReflection}, this phaser exhibits the following benefits: 1)~smaller size, due to the replacement of half-wavelength resonators by quarter-wavelength resonators, 2)~easier fabrication due to smaller required coupled-line section couplings, 3)~suppressed even-order harmonics, that are commonly observed in quarter-wavelength filters~\cite{10DIRECT-COUPLED}. Moreover, a complete design technique is proposed for this phaser.

\section{Design Methodology}
\begin{figure}[!t]
  \center
  \psfrag{0}[c][c][0.5]{\footnotesize $X_{n+1}$($\omega$)}
  \psfrag{1}[c][c][0.9]{\footnotesize (a)}
  \psfrag{2}[c][c][0.9]{\footnotesize (b)}
  \psfrag{3}[c][c][0.9]{\footnotesize (c)}
  \psfrag{4}[c][c][0.7]{\footnotesize $\lambda$/4}
  \psfrag{5}[c][c][0.7]{\footnotesize $Y_{0}$}
  \psfrag{6}[c][c][0.7]{\footnotesize $J_{0,1}$}
  \psfrag{7}[c][c][0.6]{\footnotesize $X_{1}$($\omega$)}
  \psfrag{8}[c][c][0.7]{\footnotesize $K_{0}$}
  \psfrag{9}[c][c][0.7]{\footnotesize $K_{0,1}$}
  \psfrag{s}[c][c][0.7]{\footnotesize $Z_{0}$}
  \psfrag{a}[c][c][0.7]{\footnotesize $K_{1,2}$}
  \psfrag{b}[c][c][0.7]{\footnotesize $J_{2,3}$}
  \psfrag{c}[c][c][0.55]{\footnotesize $K_{n,n+1}$}
  \psfrag{d}[c][c][0.7]{\footnotesize $J_{1,2}$}
  \psfrag{e}[c][c][0.6]{\footnotesize $K_{n,n+1}$}
  \psfrag{f}[c][c][0.5]{\footnotesize $B_{n}$($\omega$)}
  \psfrag{g}[c][c][0.6]{\footnotesize $X_{2}$($\omega$)}
  \psfrag{h}[c][c][0.7]{\footnotesize $K_{2,3}$}
  \psfrag{i}[c][c][0.5]{\footnotesize $X_{n}$($\omega$)}
  \psfrag{j}[c][c][0.5]{\footnotesize $B_{1}$($\omega$)}
  \includegraphics[width=8.5cm]{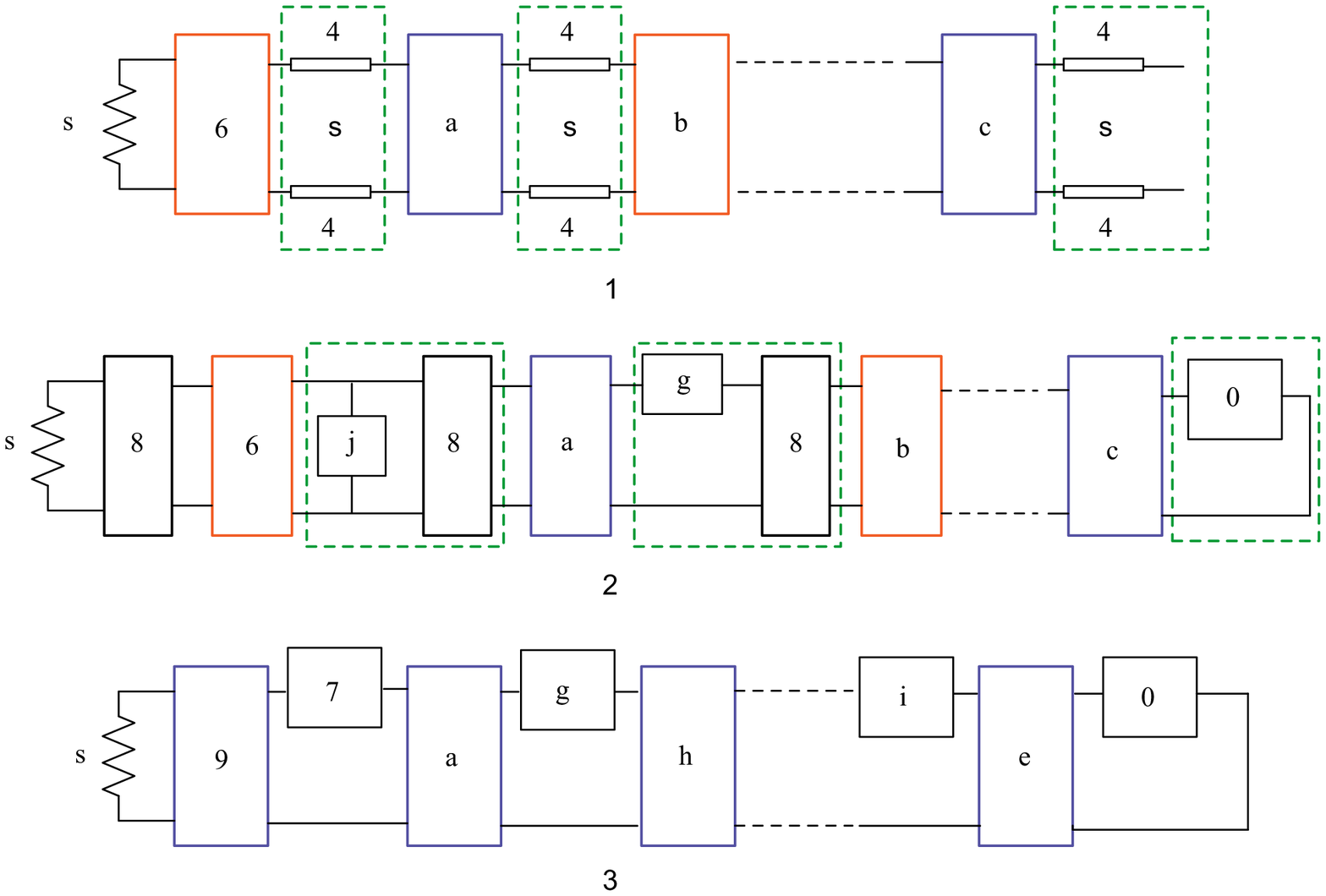}\\
  \caption{Network configuration and modeling of the phaser. (a)~Circuit model with alternating K/J-inverters and quarter-wavelength transmission lines. (b)~Equivalent network. (c) Transformed network with K-inverters.}\label{fig:quarter}
\end{figure}

Fig.~\ref{fig:quarter}(a) shows the network configuration of the proposed phaser, which is composed of quarter-wavelength transmission lines
interconnected by alternating K- and J-inverters. For simplicity, the order is assumed even, the odd case being derived in a similar fashion. The design methodology consists in transforming the network of Fig.~\ref{fig:quarter}(a) through the equivalent network of Fig.~\ref{fig:quarter}(b) into Fig.~\ref{fig:quarter}(c), which has the same form as that in~\cite{12SynthesisofNarrowbandReflection} and may therefore use the synthesis technique of~\cite{12SynthesisofNarrowbandReflection}. Here, we use microstrip technology as the implementation of Fig.~\ref{fig:quarter}(a). The design method can be also extended to other technologies.

\subsection{Circuit Transformation}

Each quarter-wavelength transmission line in Fig.~\ref{fig:quarter}(a) can be modeled by a reactance or a susceptance cascaded with a $K$-inverter~\cite{11SynthesisofWide-bandBandpassFilters}, leading to the equivalent network shown in Fig.~\ref{fig:quarter}(b). The ABCD matrix of the inverter is
\begin{equation}\label{equa:transfermatrix}
\left[\begin{array}{cc}
0 & jK_{0}\\
\dfrac{j}{K_{0}} &0
\end{array}\right],\quad\text{with}\quad K_0=Z_0,
\end{equation}
\noindent where $Z_0$ is the characteristic impedance of the transmission line, while the reactance and susceptance are~\cite{10DIRECT-COUPLED,11SynthesisofWide-bandBandpassFilters}
\begin{subequations}
\begin{align}
&\dfrac{B_{2i+1}(\omega)}{Y_{0}}=-j\cot\left(\dfrac{\pi}{2}\frac{\omega}{\omega_{0}}\right),\\
&\dfrac{X_{2i+2}(\omega)}{Z_{0}}=-j\cot\left(\dfrac{\pi}{2}\frac{\omega}{\omega_{0}}\right),
\end{align}\label{equa:XB}
\end{subequations}
\noindent where $i=0,2,\ldots,(n-1)/2$, and $Y_{0}=1/Z_{0}$ is the characteristics admittance of the line.

\begin{figure}[!t]
  \center
  \psfrag{1}[c][c][0.7]{\footnotesize $K_{0}$}
  \psfrag{4}[c][c][0.7]{\footnotesize $K_{2i,2i+1}$}
  \psfrag{3}[c][c][0.7]{\footnotesize $X_{2i+1}$}
  \psfrag{2}[c][c][0.7]{\footnotesize $J_{2i,2i+1}$}
  \psfrag{5}[c][c][0.7]{\footnotesize $B_{2i+1}$}
  \includegraphics[width=8cm]{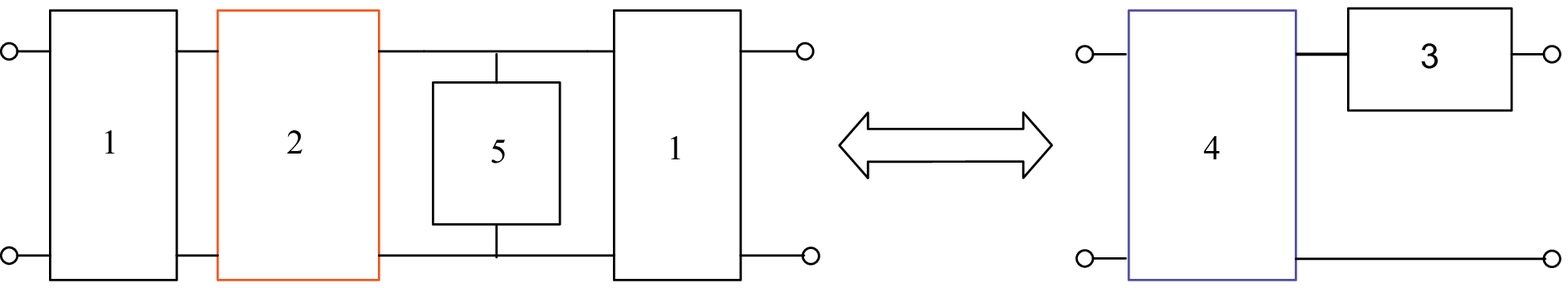}\\
  \caption{Equivalent transformation between J-inverters and K-inverters.}\label{fig:KtoJ}
\end{figure}

\begin{figure}[!t]
  \center
 \psfrag{a}[l][c][0.7]{\footnotesize $\lambda$/4 resonators}
  \psfrag{b}[l][c][0.7]{\footnotesize $\lambda$/2 resonators}
  \psfrag{c}[c][c]{\footnotesize
  $X_{\lambda/4}/Z_0=-\cot\left(\dfrac{\pi}{2}\dfrac{\omega}{\omega_{0}}\right)$
}
  \psfrag{d}[c][c]{\footnotesize $X_{\lambda/2}/Z_0=\tan\left(\pi\dfrac{\omega}{\omega_{0}}\right)$}
  \psfrag{e}[c][c]{\footnotesize Freq (GHz)}
  \psfrag{f}[c][c]{\footnotesize Normalized Reactance}
  \includegraphics[width=6cm]{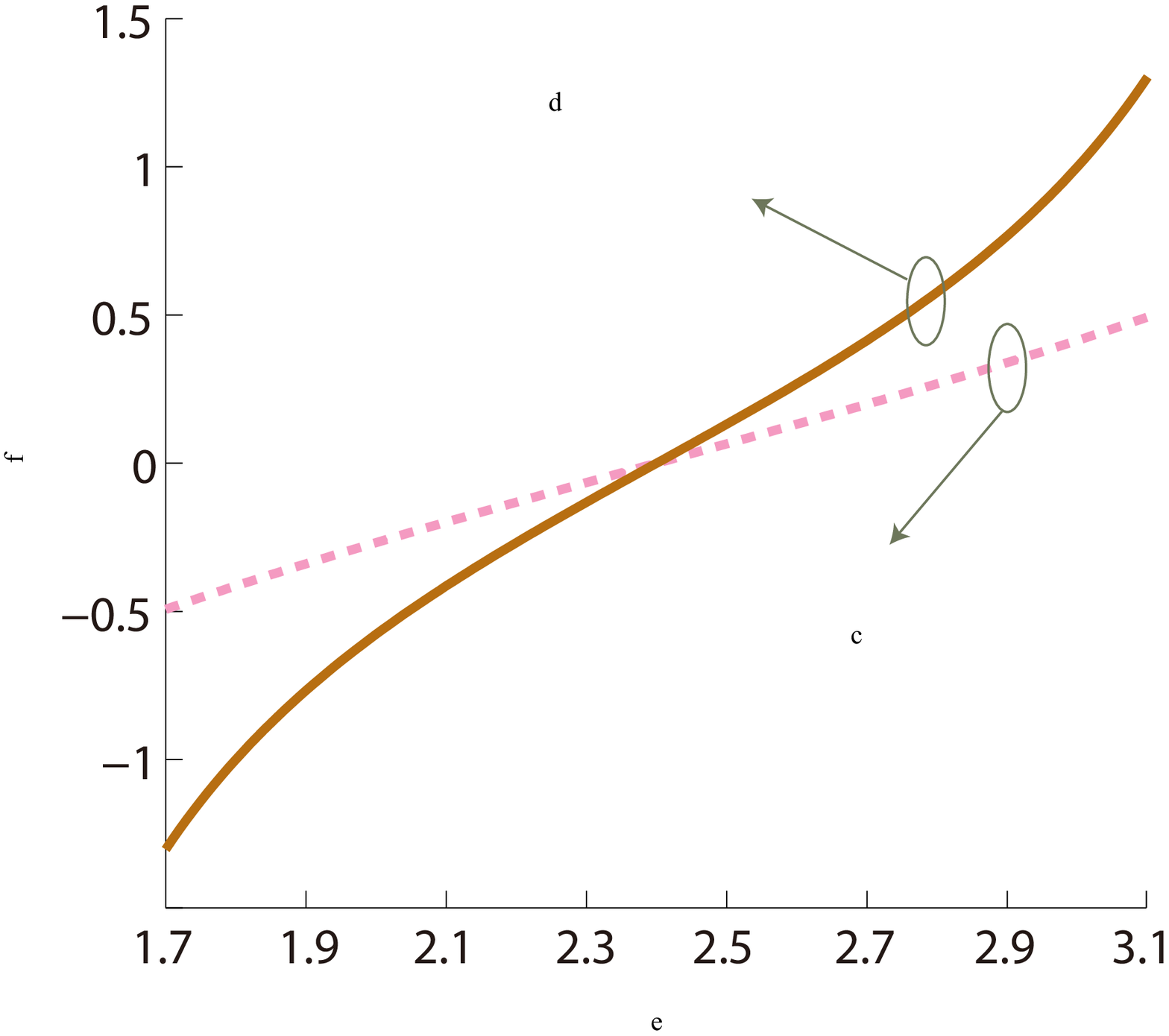}\\
  \caption{Frequency dependence of the normalized reactance for both $\lambda$/4 and $\lambda$/2 resonators.}\label{fig:impedance}
\end{figure}

The network of Fig.~\ref{fig:quarter}(b) further transforms into that of Fig.~\ref{fig:quarter}(c) upon using the transformation shown in Fig.~\ref{fig:KtoJ} which, using $K_0=Z_0$, involves the equivalences
\begin{subequations}
\begin{align}
&\dfrac{K_{2i,2i+1}}{Z_{0}}=\dfrac{J_{2i,2i+1}}{Y_{0}},\label{equa:KandJ1}\\
&\dfrac{X_{2i+1}}{Z_{0}}=\dfrac{B_{2i+1}}{Y_{0}}.
\end{align}\label{equa:KandJ}
\end{subequations}
The network in Fig.~\ref{fig:quarter}(c) exhibits the same form as that in~\cite{12SynthesisofNarrowbandReflection}, except that the reactances, given by~\eqref{equa:XB}, are different. Fig.~\ref{fig:impedance} compares these reactances. Note that the reactance slope of the quarter-wavelength transmission line is about half that of the half-wavelength transmission line. This will lead to reduced coupling coefficients, since the coupling coefficients are proportional to the reactance slope~\cite{14Hongjia-sheng}, which significantly reduces design limitations due to fabrication constraints.

The aforementioned reactance slope difference also leads to a modified mapping function between the lowpass frequency ($\Omega$) and the bandpass frequency ($\omega$) within the specified range $[\omega_0,\omega_n]$, namely
\begin{equation}\label{equa:Mapping function}
\Omega={{\cot \left( {\frac{\pi }{2}\frac{\omega }{{{\omega _0}}}} \right)} \mathord{\left/
 {\vphantom {{\cot \left( {\frac{\pi }{2}\frac{\omega }{{{\omega _0}}}} \right)} {\cot \left( {\frac{\pi }{2}\frac{{{\omega _n}}}{{{\omega _0}}}} \right)}}} \right.
 \kern-\nulldelimiterspace} {\cot \left( {\frac{\pi }{2}\frac{{{\omega _n}}}{{{\omega _0}}}} \right)}},
\end{equation}
\noindent in applying the synthesis technique of~\cite{12SynthesisofNarrowbandReflection}. Once the synthesis technique has been applied, the K-inverter values in Fig.~\ref{fig:quarter}(c) are obtained. Then the K- and J-inverter values in Fig.~\ref{fig:quarter}(a) are computed using~\eqref{equa:KandJ}.

\subsection{Inverter Implementation}

\begin{figure}[!t]
  \centering
  \psfrag{1}[c][c]{\footnotesize Via-hole}
  \psfrag{2}[c][c][1]{\footnotesize (a)}
  \psfrag{3}[c][c][1]{\footnotesize (b)}
  \psfrag{4}[c][c][1]{\footnotesize $\theta_{J}$}
  \psfrag{5}[c][c][1]{\footnotesize $\theta_{K}$}
  \psfrag{J}[c][c]{\footnotesize $J$}
  \psfrag{K}[c][c]{\footnotesize $K$}
  \includegraphics[width=3in]{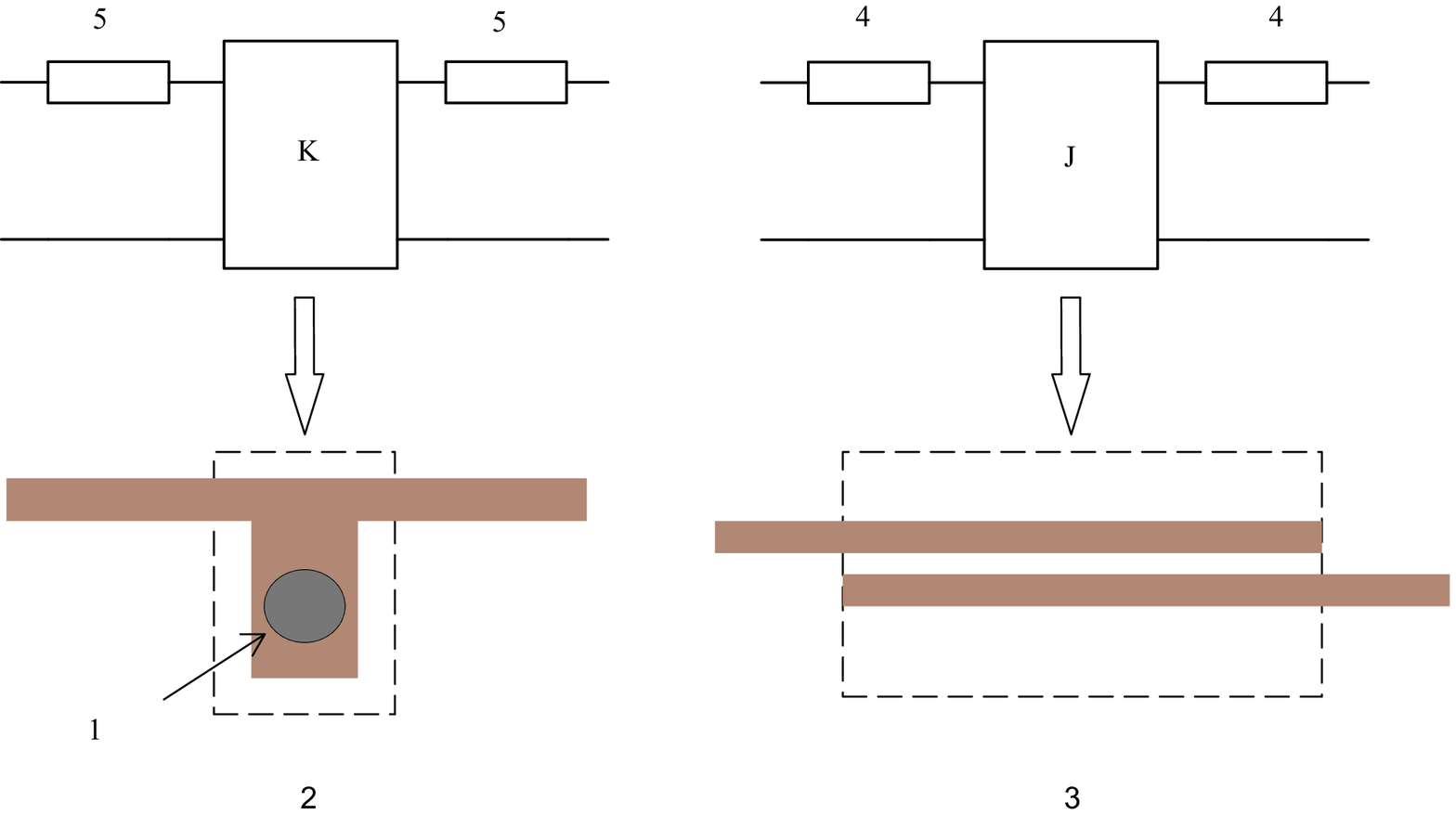}\\
  \caption{Microstrip implementation of (a)~a K-inverter and (b)~a J-inverter.}\label{fig:KandJ}
\end{figure}

A K-inverter and a J-inverter may be implemented by a metalized via hole and an open coupled-line section, respectively~\cite{13SynthesisMethodforEven-Order}, as shown in Fig.~\ref{fig:KandJ}.
To relate the circuit model to the physical structure in Fig.~\ref{fig:KandJ}, one first writes the reflection coefficient for the circuit model,
  \begin{equation}\label{equa:S11KandABCD}
  S_{11}^\text{network}=\dfrac{K^2-Z_0^2}{K^2+Z_0^2}e^{-j2\theta_{K}},
  \end{equation}
and then equates it with the full-wave response of the physical structure, i.e. $S_{11}^\text{network}=S_{11}^\text{full-wave}$, which leads to
  \begin{subequations}\label{equa:normalK}
  \begin{align}
  &|S_{11}^\text{full-wave}|=\dfrac{Z_0^2-K^2}{Z_0^2+K^2},\label{equa:normalK1}\\
  &\angle S_{11}^\text{full-wave}=\pi-2\theta_{K},\label{equa:normalK2}
  \end{align}
  \end{subequations}
  \noindent under the condition that $K/Z_0$, corresponding to the coupling level of the physical structure, be less than one. The relations for J-inverters are obtained in a similar manner and read
  \begin{subequations}\label{equa:normalJ}
  \begin{align}
  &|S_{11}^\text{full-wave}|=\dfrac{Y_0^2-J^2}{Y_0^2+J^2},\label{equa:normalJ1}\\
  &\angle S_{11}^\text{full-wave}=-2\theta_{J}.\label{equa:normalJ2}
  \end{align}
  \end{subequations}

\begin{figure}[!t]
  \centering
  \psfrag{a}[c][c][0.85]{\footnotesize \shortstack{specification\\$\tau(\omega)$}}
  \psfrag{b}[c][c][0.85]{\footnotesize \shortstack{synthesis using~\cite{12SynthesisofNarrowbandReflection} and\\ mapping function \eqref{equa:Mapping function}}}
  \psfrag{c}[c][c][0.85]{\footnotesize \shortstack{K-inverters\\in Fig.~\ref{fig:quarter}(c)}}
  \psfrag{d}[c][c][0.85]{\footnotesize \shortstack{K/J-inverters\\in Fig.~\ref{fig:quarter}(a)}}
  \psfrag{e}[c][c][0.85]{\footnotesize \shortstack{implementation\\using~\eqref{equa:normalK},~\eqref{equa:normalJ}, and\\full-wave analysis}}
  \psfrag{f}[c][c][0.85]{\footnotesize converge?}
  \psfrag{g}[c][c][0.85]{\footnotesize end}
  \psfrag{h}[c][c][0.85]{\footnotesize yes}
  \psfrag{i}[c][c][0.85]{\footnotesize no}
  \psfrag{j}[c][c][0.85]{\footnotesize distortion correction~\cite{12SynthesisofNarrowbandReflection}}
  \psfrag{k}[c][c][0.85]{\footnotesize \shortstack{applying\\eq. \eqref{equa:KandJ1}}}
  \includegraphics[width=8.6cm]{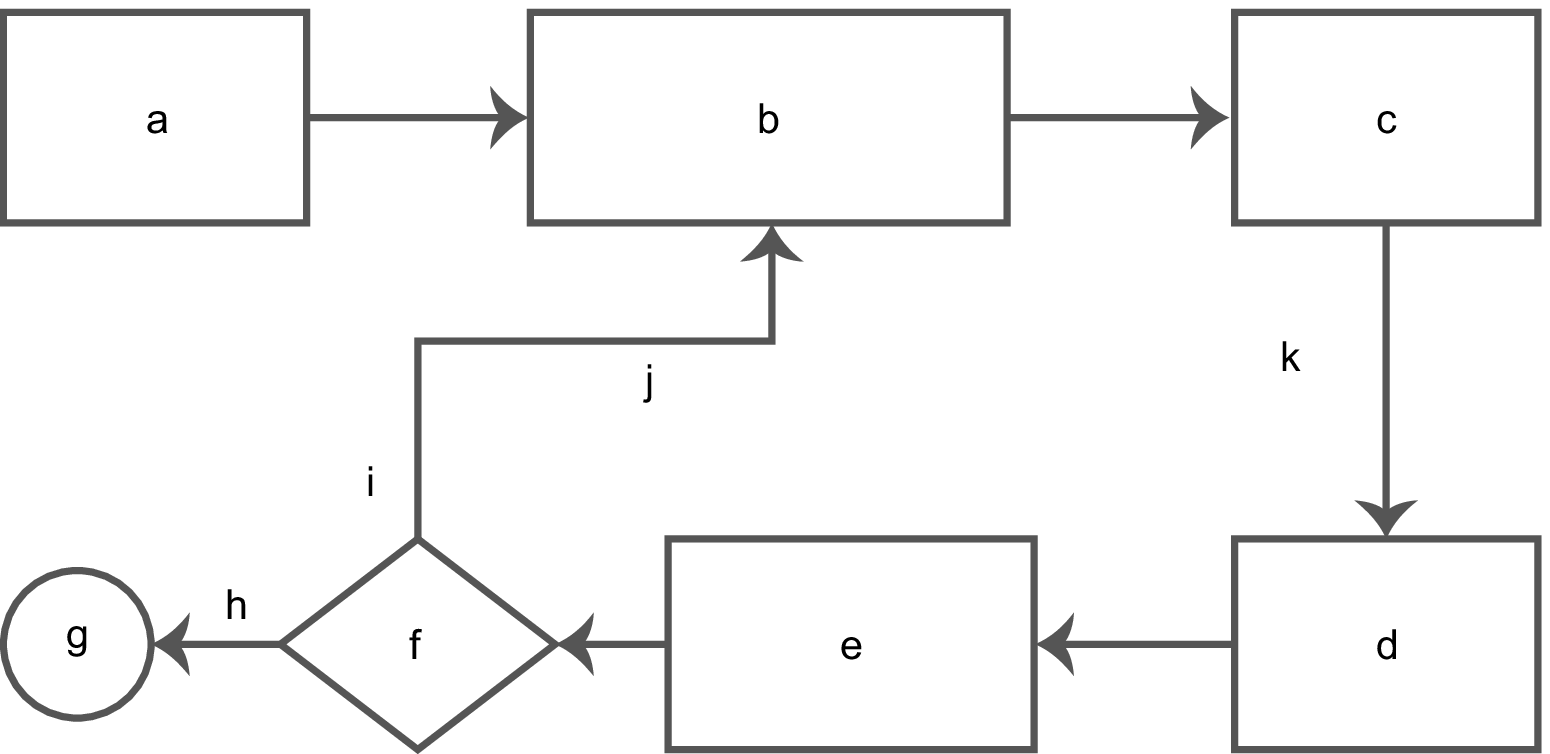}\\
  \caption{Synthesis procedure for the design of quarter-wavelength phasers.}\label{fig:flowchart}
\end{figure}

To apply the above equations, one first obtains the required K- and J-inverter values, then uses~\eqref{equa:normalK1} and \eqref{equa:normalJ1} to calculate the required $|S_{11}^\text{full-wave}|$, and finally performs a full-wave analysis to align the magnitude response with $|S_{11}^\text{full-wave}|$. Once this has been done, one calculates $\theta_K$ and $\theta_J$ using the phase response of the physical structure using~\eqref{equa:normalK2} and \eqref{equa:normalJ2}. A flowchart of the whole design procedure is presented in Fig.~\ref{fig:flowchart}.

\section{Example}

\begin{figure}[!t]
  \centering
  \psfrag{a}[c][c][0.9]{\footnotesize \shortstack{quarter-wavelength\\phaser}}
  \psfrag{b}[l][c][0.9]{\footnotesize \shortstack{half-wavelength\\phaser}}
  \psfrag{c}[c][c]{\footnotesize Freq (GHz)}
  \psfrag{d}[c][c]{\footnotesize Groupdelay (ns)}
  \includegraphics[width=3.3in]{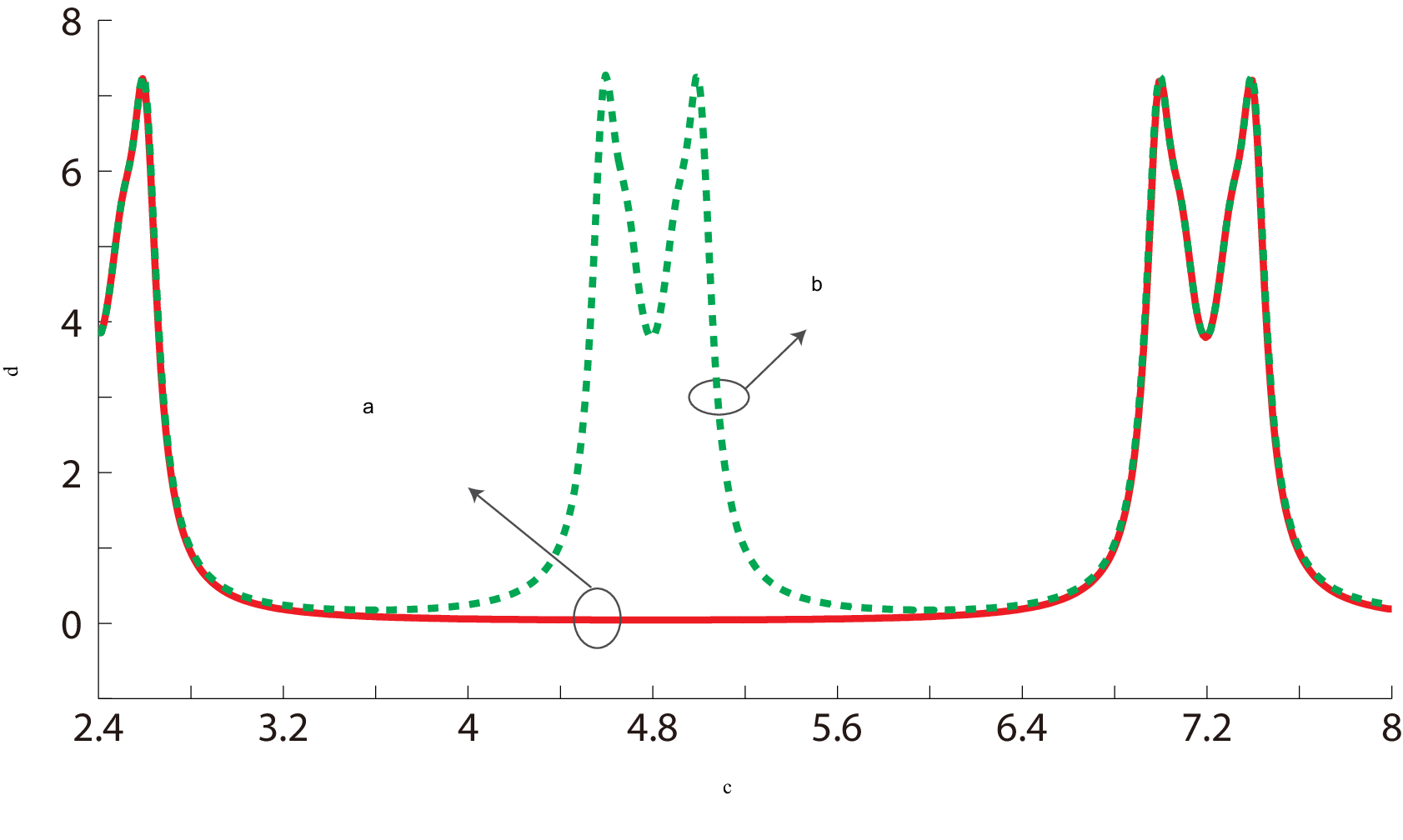}\\
  \caption{Compared group delay responses for a quarter-wavelength phaser and a half-wavelength phaser.}\label{fig:groupdelay}
\end{figure}

To validate the proposed design method, a quarter-wavelength phaser with linear group delay within the frequency band $2.4-2.6$ GHz is design. In addition, a half-wavelength phaser with the same specifications is also designed for comparison. The two results are shown in~Fig.~\ref{fig:groupdelay}. Note that both exhibit the same response in the specified frequency band (and its odd multiple), but differ around the second (and naturally other even) harmonic bands. So, the quarter-wavelength phaser conveniently suppresses the even-order harmonics\footnote{This could be predicted by the fact that the transmission line sections in Fig.~\ref{fig:quarter} are $\lambda/2$ long at the second harmonic, so that they become transparent (one turn in the Smith chart), leaving only (non-resonant) inverters in the network.} while the half-wavelength phaser does not. Table~\ref{table_example} shows the normalized inverter values (coupling coefficients) of the two phasers. Note that quarter-wavelength phaser's values are about half those of the half-wavelength phaser.

\begin{figure}[!t]
  \centering
  \psfrag{a}[c][c][1]{\footnotesize phaser with $\lambda$/4 resonators}
  \psfrag{b}[c][c][1]{\footnotesize phaser with $\lambda$/2 resonators}
   \psfrag{c}[c][c][1]{\footnotesize Unit: mm}
  \includegraphics[width=3.3in]{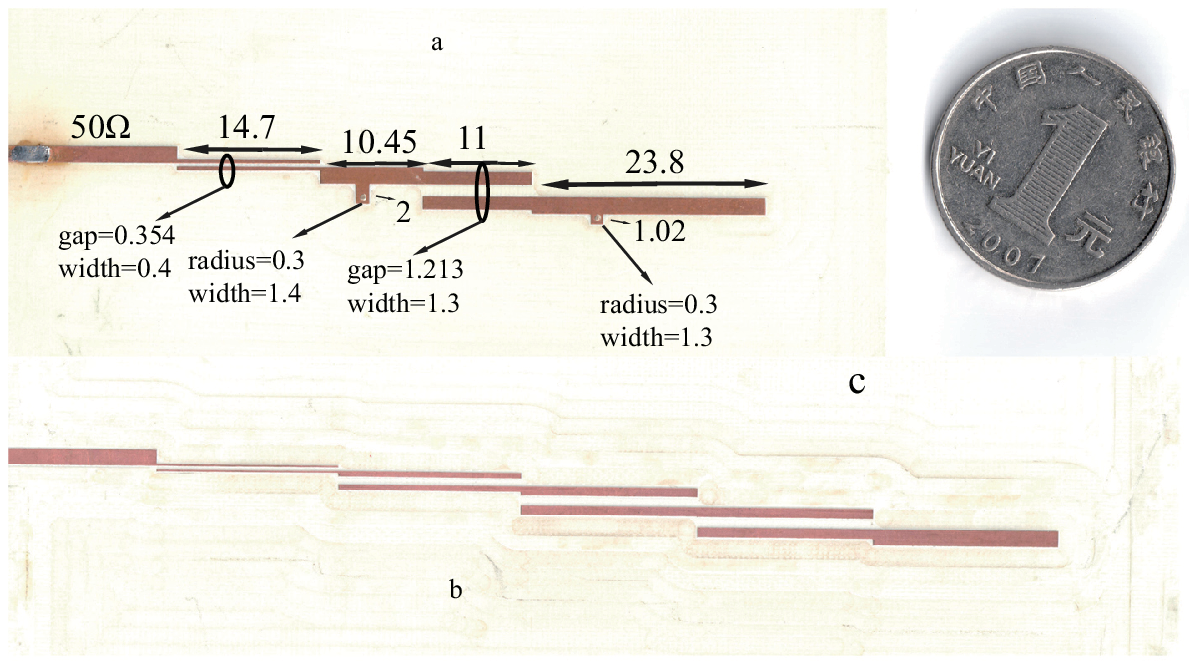}\\
  \caption{Fabricated photograph of the designed quarter-wavelength phaser and half-wavelength phaser.}\label{fig:fabricated}
\end{figure}

\begin{figure}[!t]
\centering
  \psfrag{a}[l][c]{\footnotesize Simulated}
  \psfrag{b}[l][c]{\footnotesize Prescribed}
  \psfrag{c}[l][c]{\footnotesize Measured}
  \psfrag{g}[c][c]{\footnotesize Groupdelay (ns)}
  \psfrag{f}[c][c]{\footnotesize Freq (GHz)}
  \psfrag{s}[c][c]{\footnotesize $|$$S_{11}$$|$ (dB)}
  \includegraphics[width=3.3in]{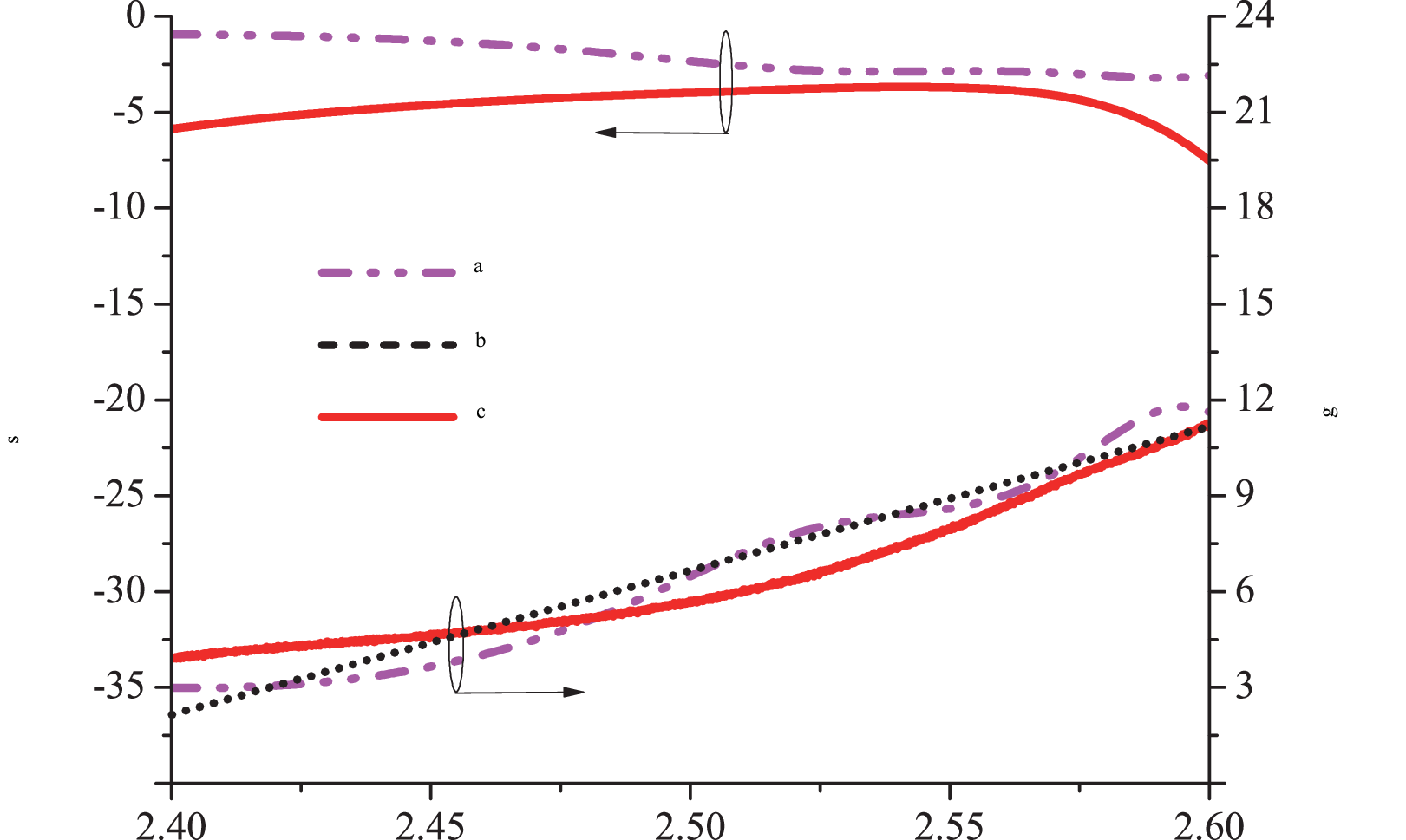}\\
  \caption{The simulated and measured return loss and group delay of the phaser.}\label{fig:s_and_group}
\end{figure}

A $4^\text{th}$-order microstrip prototype, shown in Fig.~\ref{fig:fabricated}, has been fabricated, on a $0.762$ mm-thick Rogers RO4350 substrate ($\epsilon_r=3.66, \tan\delta=0.004$). The smaller size of quarter-wavelength phaser is clearly apparent. Fig.~\ref{fig:s_and_group} shows the simulation and measurement results. The measured group delay follows the simulated and specified ones, while the magnitude shows a deviation due to fabrication tolerance.

\begin{table}[!t]
\center
\renewcommand{\arraystretch}{1.3}
\caption{Normalized Inverter Values ($K/Z_0$ and $J/Y_0$)}
\label{table_example}
\centering
\begin{tabular}{c|c|c|c|c}
  \hline
  \hline
  &1&2&3&4\\
  \hline
  half-wavelength phaser& 0.6207 & 0.3035 & 0.1832 & 0.1790 \\
  \hline
  quarter-wavelength phaser& 0.4484 & 0.1519 & 0.0893  & 0.0869 \\
  \hline
  \hline
\end{tabular}
\end{table}

\section{Conclusion}
A compact reflection-type phaser using quarter-wavelength transmission line resonators has been proposed. The design method was presented and validated by an illustrative example. It has been shown that the proposed phaser exhibits the benefits of smaller size, easier fabrication and suppressed even-order harmonics compared with the half-wavelength phaser designed using~\cite{12SynthesisofNarrowbandReflection}.

\bibliographystyle{IEEEtran}
\bibliography{mybib}

\end{document}